\documentclass[12pt,twoside]{article}
\usepackage{fleqn,espcrc1}
\usepackage{graphicx}

\title{Low-$Q^{2}$ low-$x$ Structure Function Analysis of CCFR data for $F_{2}$}

\author{B.~H.~Tamminga \address{Columbia University, New York, NY
10027},T.~Adams \address{Kansas State University, Manhattan, KS
66506}, A.~Alton$^{b}$, C.~G.~Arroyo$^{a}$, S.~Avvakumov
\address{University of Rochester, Rochester, NY 14627}, L.~de~Barbaro
\address{Northwestern University, Evanston, IL 60208}, P.~de~Barbaro
$^{c}$, A.~O.~Bazarko$^{a}$, R.~H.~Bernstein \address{Fermi National
Accelerator Laboratory, Batavia, IL 60510}, A.~Bodek$^{c}$, T.~Bolton
$^{b}$, J.~Brau \address{University of Oregon, Eugene, OR 97403},
D.~Buchholz$^{d}$, H.~Budd$^{c}$, L.~Bugel$^{e}$, J.~Conrad$^{a}$,
R.~B.~Drucker$^{f}$, J.~A.~Formaggio$^{a}$, R.~Frey$^{f}$,
J.~Goldman$^{b}$, M.~Goncharov$^{b}$, D.~A.~Harris$^{c}$,
R.~A.~Johnson \address{University of Cincinnati, Cincinnati, OH
45221}, J.~H.~Kim$^{a}$, B.~J.~King$^{a}$, T.~Kinnel
\address{University of Wisconsin, Madison, WI 53706},
S.~Koutsoliotas$^{a}$, M.~J.~Lamm$^{e}$, W.~Marsh$^{e}$, D.~Mason$^{f}$,
K.~S.~McFarland$^{c}$, C.~McNulty$^{a}$, S.~R.~Mishra$^{a}$,
D.~Naples$^{b}$, P.~Nienaber$^{e}$, A.~Romosan$^{a}$, W.~K.~Sakumoto$^{c}$,
H. Schellman$^{d}$, F.~J.~Sciulli$^{a}$, W.~G.~Seligman$^{a}$,
M.~H.~Shaevitz$^{a}$, W.~H.~Smith$^{h}$, P.~Spentzouris$^{a}$,
E.~G.~Stern$^{a}$, M.~Vakili$^{g}$, A.~Vaitaitis$^{a}$, V.~Wu$^{g}$,
U.~K.~Yang$^{c}$, J.~Yu$^{e}$, G.~P.~Zeller$^{d}$, E.~D.~Zimmerman$^{a}$}

\begin{document}
\maketitle

\begin{abstract}
Analyses of structure functions (SFs) from neutrino and muon deep inelastic
scattering (DIS) data have shown discrepancies in F$_{2}$ for $x
< 0.1$.  A new SF analysis of the CCFR collaboration
data examining regions in $x$ down to $x=.0015$ and $0.4 < Q^{2} < 1.0$
is presented.  Comparison to corrected charged lepton scattering
results for $F_{2}$ from the NMC and E665 experiments are
made. Differences between $\mu$ and $\nu$ scattering allow that the
behavior of $F_{2}^{\mu}$ could be different from $F_{2}^{\nu}$ as
$Q^{2}$ approaches zero.  Comparisons between $F_{2}^{\mu}$ and
$F_{2}^{\nu}$ are made in this limit.
\end{abstract}

High-energy neutrinos are a unique probe for understanding the parton
properties of nucleon structure. Combinations of $\nu$ and
$\overline{\nu }$ DIS data are used to determine the $F_{2}$ and $xF_{3}$
SFs which determine the valence, sea, and gluon parton distributions
in the nucleon \cite{GRVMRS,CTEQ}. The universalities of parton
distributions can also be studied by comparing neutrino and charged
lepton scattering data. Past measurements have indicated that
$F_{2}^{\nu }$ differs from $F_{2}^{e/\mu }$ by 10-15\% in the low-$x$
region \cite{seligman:1997mc}.  These differences are larger than the
quoted combined statistical and systematic errors of the measurements
and may indicate the need for modifications of the theoretical
modeling to include higher-order or new physics contributions. We
present a new analysis of the CCFR collaboration $\nu$-$N$ DIS data
in a previously unexplored kinematic region.  In this low-$x$ and
low-$Q^{2}$ region, the discrepancy between $F_{2}^{\nu}$ and $F_{2}^{\mu}$
persists.  However, in this kinematic region some
differences in $F_{2}$ from neutrino and charged lepton data may result
from differences in the properties of weak and electromagnetic
interactions.  Within the PCAC nature of $\nu$-$N$ DIS,
$F_{2}^{\nu}$ should approach a constant as $Q^{2}$ approaches zero,
while $F_{2}^{e/\mu}$ for charged lepton DIS should approach zero. A
determination of this constant is presented.

The $\nu$ DIS data were taken in two high-energy high-statistics
runs, FNAL E744 and E770, in the Fermilab Tevatron fixed-target
quadrupole triplet beam (QTB) line by the CCFR collaboration. The
detector, described in Refs.~\cite{hadcal,mucal}, consists of a target
calorimeter instrumented with both scintillators and drift chambers
for measuring the energy of the hadron shower $E_{HAD}$ and the $\mu$
angle $\theta _{\mu }$, followed by a toroid spectrometer for
measuring the $\mu$ momentum $p_{\mu }$. There are 950,000 $\nu _{\mu
}$ events and 170,000 $\overline{\nu }_{\mu }$ events in the data
sample after fiducial-volume cuts, geometric cuts, and kinematic cuts
of $p_{\mu }$ $>15\ $ GeV, $\theta _{\mu }<150\ $ mr, $E_{HAD}$ $>10\
GeV$, and $30<E_{\nu }<360\ GeV$, to select regions of high efficiency
and small systematic errors in reconstruction.

In order to extract the SFs from the number of observed $\nu _{\mu }$
and $\overline{\nu }_{\mu }$ events, determination of the flux was
neccesary~\cite{WGSthesis,Auc,BelRein}.  The
cross-sections, multiplied by the flux, are compared to the observed
number of $\nu $-$N$ and $\overline{\nu }$-$N$ events in each $x$ and
$Q^{2}$ bin to extract $F_{2}(x,Q^{2})$ and $xF_{3}(x,Q^{2})$.
Determination of muon and hadron energy calibrations from the previous
CCFR analysis were used in the present analysis.  These calibrations
were determined from test beam data collected during the course of the
experiment \cite{hadcal,mucal}. Changes in the SF extraction to extend
the analysis into the low-$Q^{2}$, low-$x$ region include incorporation
of an appropriate model below $Q^{2}$ of 1.35 GeV$^{2}$, in this case
we chose the GRV \cite{GRVMRS} model of PDFs.  The data have been
corrected using the leading order Buras-Gaemers model \cite{BG} for slow
rescaling \cite{Bar76,TMC}, with charm mass of 1.3 GeV and for the
difference in $xF_{3}^{\nu}-xF_{3}^{\nu}$. In addition, corrections
for radiative effects \cite{radcorr}, non-isoscalarity of the Fe
target, and the mass of the $W$-boson propagator were applied.  Due to
the systematic uncertainty in the model, the radiative correction
error dominates in the lowest $x$ bins. Other significant systematics
across the entire kinematic region include the value of $R$, which
comes from a global fit to the world's measurements \cite {Rworld}.

\begin{figure}[tbp]
\centerline{\includegraphics*[scale=0.7]{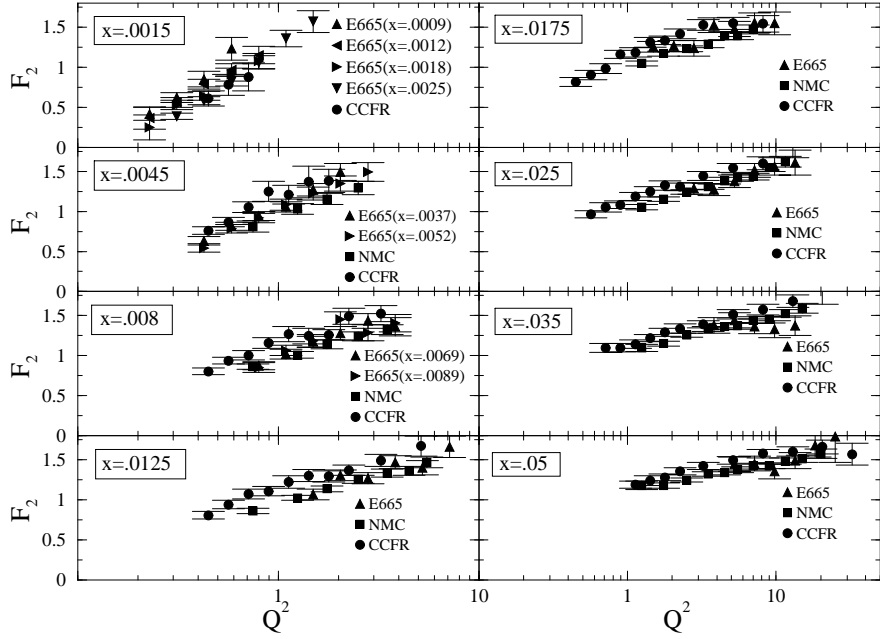}}
\caption{$F_{2}$ from CCFR $\nu $-Fe DIS compared to $F_{2}$ from $\mu
$D DIS. Errors bars are statistical and systematic added in
quadrature. The $\mu$ data have been corrected as described in the
text.}
\label{fig:compare}
\end{figure}

The SF $F_{2}$ from $\nu$ DIS on iron can be compared to $F_{2}$ from
charged lepton DIS on isoscalar targets. To make this comparison, two
corrections must be made to the charged lepton data. For deuterium
data, a heavy nuclear target correction must be made to convert
$F_{2}^{\ell D}$ to $F_{2}^{\ell \,Fe}$ \cite{heavytarget}. Second, a
correction was made to account for the different quark charge involved
in the charged lepton DIS interactions \cite{seligman:1997mc}.  The
errors on the nuclear and charge corrections are small compared to the
statistical and systematic errors on both the CCFR and NMC data. The
corrected SF, $F_{2}$, from $\mu$ DIS experiments NMC and
E665\cite{NMC,F2l} along with CCFR for lowest $x$-bins is shown in
Fig.~\ref{fig:compare}. The new analysis allows comparison to E665
data, which is in the low-$x$, low-$Q^{2}$ region. Error bars for CCFR
and E665 data are large in the x-bin, x=.0015.  However, In the next
x-bin, $x=.0045$, there is clearly as much as a 20\% discrepancy
between the NMC $F_{2}^{\mu}$ and the CCFR $F_{2}^{\nu}$ and an
approximately 10\% discrepancy between CCFR and E665.  As the value of
$x$ increases, the discrepancy decreases; there is agreement between
CCFR and the charged lepton experiments above $x=0.1$.

The discrepancy between CCFR and NMC at low-$x$ is outside the
experimental systematic errors quoted by the groups.  Several
suggestions for an explanation have been put forward. One suggestion
\cite{CTEQ1}, that the discrepancy can be entirely explained by a
large strange sea, is excluded by the CCFR dimuon analysis which
directly measures the strange sea \cite{AOB}.  Another is that the
strange sea may not be the same as the anti-strange sea distribution.
Data from both NMC and CCFR do not support this
possibility \cite{ssbar}. Another possibilty is that the heavy nuclear
target correction may be different between neutrinos and charged leptons.
Heavy target corrections used in this paper are determined by NMC for
charged lepton-nucleon DIS data and applied to NMC and E665 only; no
charged lepton correction data is applied to $\nu$ data.  Another
possibility that has been proposed would have a large symmetry
violation in the sea quark \cite{Boros:1998zd}, but recently the
model has been ruled out by the CDF $W$ charge asymmetry measurements
\cite{CDF}.  Finally, in the low-$x$ and low-$Q^{2}$ region, some of
the discrepancy may be accounted for by the differences in behavior of
$F_{2}$ as $Q^{2}$ approaches zero, although this can only address the
$x < 0.0175$ region.

In charged lepton DIS, the SF, $F_{2}$, is constrained by gauge
invariance to vanish linearly with $Q^{2}$ at $Q^{2}=0$.  Donnachie
and Landshoff predict that in the low-$Q^{2}$ region, $F_{2}^{\mu}$
will follow the form \cite{DL} $C \left( \frac{Q^{2}}{Q^{2}+A^{2}}
\right)$.  However, in the case of neutrino DIS, the PCAC nature
of the weak interaction contributes a nonzero component to $F_{2}$ as
$Q^{2}$ approaches zero.  Donnachie and Landshoff predict that
$F_{2}^{\nu}$ should follow a form with a non-zero contribution at
$Q^{2}=0$: $\frac{C}{2} \left( \frac{Q^{2}}{Q^{2}+A^{2}} + \frac{Q^{2}
+ D}{Q^{2} + B^{2}} \right) $.  Using NMC data we fit to the form
predicted for $e/\mu$ DIS, extracting the parameter A.  Inserting
this value for A into the form predicted for $\nu$ DIS, we fit CCFR data to
extract parameters B,C,D, and determine the value of $F_{2}$ at
$Q^{2}=0$. Only data below $Q^{2}=1.35$ GeV$^{2}$ are used in the
fits.  The CCFR x-bins having enough data for a good fit in this
$Q^{2}$ region are $x=.0045$, $x=.0080$, $x=.0125$, $x=.0175$.  Table
\ref{tab:fitresults} shows the results of the fits.  The values of
$F_{2}$ at $Q^{2}$=$0$ in the three highest $x$-bins are statistically significant
and in agreement with each other.  The lowest x-bin is consistent with
the other results.

\begin{table}[tbp]
\caption{Fit results for NMC and CCFR data. NMC is fit to Eq. 3 and
parameter A extracted:  $A = 1.00 \pm 0.17$. CCFR data is fit to Eq. 4 with A extracted
from NMC fit.  B, C, D and $F_{2}$ at $ Q^{2}$ results shown below.}
\begin{center}
\begin{tabular}{cccccc}
$x$  & $B$ & $C$ & $D$ & $F_{2}^{\nu}(Q^{2}=0)$ & $\chi^{2}$ \\
\hline \hline
$0.0045$ & $1.54 \pm 0.03$ & $2.57 \pm 0.29$ & $0.40 \pm 0.23$ & $0.22 \pm 0.13$ & $1.01$\\
$0.0080$ & $1.51 \pm 0.04$ & $2.34 \pm 0.06$ & $0.64 \pm 0.06$ & $0.33 \pm 0.04$ & $1.04$\\
$0.0125$ & $1.50 \pm 0.04$ & $2.28 \pm 0.05$ & $0.70 \pm 0.06$ & $0.36 \pm 0.04$ & $0.70$\\
$0.0175$ & $1.51 \pm 0.04$ & $2.31 \pm 0.05$ & $0.64 \pm 0.06$ & $0.33 \pm 0.04$ & $0.80$\\
\end{tabular} 
\end{center}
\label{tab:fitresults}
\end{table}

In summary, a comparison of $F_{2}$ from $\nu$ DIS to that from
$\mu$ DIS continues to show good agreement above $x=0.1$ but a
difference at smaller $x$ that grows to 20\% at $x = 0.0045$. The
experimental systematic errors between the two experiments, and
improved theoretical analyses of massive charm production in both
neutrino and muon scattering are both presently being investigated as
possible reasons for this discrepancy.  Some of this low-$x$
discrepancy may be explained by the different behavior of $F_{2}$ from
$\nu$ DIS to that from $e/\mu$ DIS at $Q^{2}=0$.  CCFR $F_{2}^{\nu}$
data appear to approach a non-zero constant at $Q^{2}=0$.


\end{document}